\documentclass{aastex631}

\begin{document}

\title{Four Total Eclipsing Contact Binary Systems: The First Photometric Light Curve Solutions Employing TESS and Gaia Surveys}

\author{Atila Poro}
\altaffiliation{Corresponding author: atila.poro@obspm.fr}
\affiliation{LUX, Observatoire de Paris, CNRS, PSL, 61 Avenue de l'Observatoire, 75014 Paris, France}
\affiliation{Astronomy Department of the Raderon AI Lab., BC., Burnaby, Canada}

\author{Razieh Aliakbari}
\affiliation{Physics Society of Iran (PSI), Tehran, Iran}

\author{Hossein Azarara}
\affiliation{Physics Department, Shahid Bahonar University of Kerman, Kerman, Iran}

\author{Asma Ababafi}
\affiliation{Independent Astrophysics Researcher, Dezful, Iran}

\author{Sadegh Nasirian}
\affiliation{Thaqib Astronomical Association, Rasht, Iran}

\begin{abstract}
We presented the first photometric light curve solutions of four W Ursae Majoris (W UMa)-type contact binary systems. This investigation utilized photometric data from the Transiting Exoplanet Survey Satellite (TESS) and Gaia Data Release 3 (DR3). We used the PHysics Of Eclipsing BinariEs (PHOEBE) Python code and the Markov Chain Monte Carlo (MCMC) method for these light curve solutions. Only TIC 249064185 among the target systems needed a cold starspot to be included in the analysis. Based on the estimated mass ratios for these total eclipse systems, three of them are categorized as low mass ratio contact binary stars. The absolute parameters of the systems were estimated using the Gaia DR3 parallax method and the orbital period and semi-major axis ($P-a$) empirical relationship. We defined that TIC 318015356 and TIC 55522736 systems are A-subtypes, while TIC 249064185 and TIC 397984843 are W-subtypes, depending on each component's effective temperature and mass. We estimated the initial masses of the stars, the mass lost by the binary system, and the systems' ages. We displayed star positions in the mass-radius, mass-luminosity, and total mass-orbital angular momentum diagrams. In addition, our findings indicate a good agreement with the mass-temperature empirical parameter relationship for the primary stars.
\end{abstract}


\section{Introduction}
The W UMa-type eclipsing contact binary systems have two late-type stars with a short orbital period. The two components in the system overfill their respective critical Roche lobes and share a common envelope (\citealt{1959cbs..book.....K}). 

Contact binaries are generally classified into two A and W subtypes (\citealt{binnendijk1970orbital}). The subtype of a system can be recognized by estimating its temperatures and the mass of the stars.

The contact systems' stars are transferring mass and energy to each other (\citealt{1979ApJ...231..502L}), and their orbital period changes in this process. The orbital period of contact systems plays a role in studying empirical relationships of the fundamental parameters and is effective in the evolutionary process of these systems (\citealt{2021ApJS..254...10L}, \citealt{2022MNRAS.514.5528L}). There are studies conducted on the upper and lower cut-offs of these systems' orbital periods (\citealt{2020MNRAS.497.3493Z}). The investigations show that contact systems' orbital periods usually lie between about 0.2 and 0.6 days (\citealt{2024RAA....24a5002P}).

The asymmetric light curves in contact and near-contact binaries are commonly observed in phases 0.25 and 0.75. This phenomenon is generally referred to as the O'Connell effect (\citealt{1951MNRAS.111..642O}), which is crucial for studying a star's magnetic activity. This asymmetry in the light curves is solved with one or more starspot(s).

In recent decades, surveys such as Kepler (\citealt{2010AAS...21510101B}), the Catalina Sky Survey (CSS, \citealt{2017MNRAS.465.4678M}), the All-Sky Automated Survey for Supernovae (ASAS-SN, \citealt{2018MNRAS.477.3145J}), the Asteroid Terrestrial-impact Last Alert System (ATLAS, \citealt{2018AJ....156..241H}), and the TESS mission (\citealt{2015JATIS...1a4003R}) have contributed to a dramatic rise in the number of known eclipsing contact binaries. Our knowledge of W UMa stars is still incomplete, even with this previous improvement. Light curve modeling, mass ratio estimation, and component temperature enable the determination of the absolute stellar parameters: the masses, radii, and brightness of the components in solar units (\citealt{1999ebs..conf.....K}). Also, increasing the number of well-studied W UMa binaries with absolute parameters enables astronomers to derive empirical parameter relationships.

In this investigation, we presented a photometric analysis for four contact binary systems. The paper is organized as follows: Section 2 provides specifications of the target systems and the dataset. Section 3 contains the light curve solution for the system. Section 4 presents the estimation of the absolute parameters. Finally, Section 5 includes the discussion and conclusion.

\vspace{1cm}
\section{Target Systems and Dataset}
For this investigation, four contact binary stars were chosen, including TIC 249064185, TIC 318015356, TIC 397984843, and TIC 55522736. According to the appearance of their light curves, these systems were total eclipses, and in-depth photometric analysis has not yet been performed on them. The target systems are all categorized as contact binary stars in several catalogs, such as the ASAS-SN (\citealt{2018MNRAS.477.3145J}), ZTF\footnote{Zwicky Transient Facility} (\citealt{2020ApJS..249...18C}), GCVS\footnote{General Catalogue of Variable Stars} (\citealt{2017ARep...61...80S}), APASS9\footnote{AAVSO Photometric All-Sky Survey} (\citealt{2015AAS...22533616H}), and VSX\footnote{Variable Star Index, \url{https://www.aavso.org/vsx/}} catalogs.
The general characteristics of the targets are presented in Table \ref{system-info} include the names of the systems in two catalogs, coordinates and distance from the Gaia DR3 database, and reference ephemeris including orbital period $P_0$ and minimum time $t_0$ from the literature. We utilized the online tool\footnote{\url {https://astroutils.astronomy.osu.edu/time/hjd2bjd.html}} to convert the Heliocentric Julian Date ($HJD$) to the Barycentric Julian Date ($BJD_{TDB}$) since $t_0$ was reported in the $HJD$.

\begin{table*}
\caption{Specifications of the target systems.}
\centering
\begin{center}
\footnotesize
\begin{tabular}{c c c c c c c}
\hline
Name & Name & RA$.^\circ$ & Dec$.^\circ$ & $d$ & $P_0$ & $t_0$\\
TIC & Gaia DR3 & (J2000) & (J2000) & (pc) & (day) & ($BJD_{TDB}$)\\
\hline
249064185 & 3208726533053539072 & 79.260986 & -5.263103 & 744(22) & 0.3501901 & 2457700.651071\\
318015356 & 2992215135718505984 & 90.045663 & -15.829143 & 644(6) & 0.4169211 & 2458037.850993\\
397984843 & 534022116032892032 & 23.410532 & 71.043707 & 1155(19) & 0.4576875 & 2458028.072765\\
55522736 & 5543014496502562432 & 126.173182 & -35.458112 & 1071(15) & 0.3897012 & 2458205.512582\\
\hline
\end{tabular}
\end{center}
\label{system-info}
\end{table*}

We used TESS data to analyze these systems. TESS has a time series and good-quality data for each target system. The TESS data was obtained from the database of the Mikulski Archive for Space Telescopes (MAST)\footnote{\url{https://mast.stsci.edu/portal/Mashup/Clients/Mast/Portal.htmL}}. TESS-style curves were extracted from the MAST using the LightKurve\footnote{\url{https://docs.lightkurve.org}} code. The data were detrended using the TESS Science Processing Operations Centre (SPOC) pipeline (\citealt{2016SPIE.9913E..3EJ}). We selected systems' light curves based on the most recent or better-quality observed TESS sector that was available.
The characteristics of the observations and sectors used in this study of TESS are listed in Table \ref{infoTESS}. The apparent magnitude ($V$) of the systems in Table \ref{infoTESS} is from the TESS Input Catalog (TIC) v8.2 database.

\begin{table*}
\renewcommand\arraystretch{1.2}
\caption{Specifications of the target systems from the TESS.}
\centering
\begin{center}
\footnotesize
\begin{tabular}{c c c c c}
\hline
System & $V$(mag.) & Sector & Exposure length(s) & Observation start\\
\hline
TIC 249064185 & 13.626(149) & 32 & 600 & 2020\\
TIC 318015356 & 11.870(22) & 33 & 600 & 2020\\
TIC 397984843 & 14.151(149) & 58 & 200 & 2022\\
TIC 55522736 & 13.728(114) & 62 & 200 & 2023\\
\hline
\end{tabular}
\end{center}
\label{infoTESS}
\end{table*}

We also used Gaia space-based telescope data for this study. Time-series photometric data were accessed through the online service provided by the Gaia team at the Astronomisches Rechen-Institut (ARI)\footnote{\url{https://gaia.ari.uni-heidelberg.de}}. The data were downloaded in VOTable format, a standard file format used in the Virtual Observatory, and analyzed using TOPCAT (\citealt{2005ASPC..347...29T}). We extracted Gaia's $G$ photometric filter data.

\vspace{1cm}
\section{Light Curve Solutions}
The light curves of the contact systems were analyzed using the PHOEBE Python code version 2.4.9 (\citealt{2016ApJS..227...29P}, \citealt{2020ApJS..250...34C}). We also made use of the BSN application\footnote{\url{https://bsnp.info/}}. The four target binary systems' light curves were investigated for the first time. The short orbital period and the type described in the catalogs led us to choose a contact mode, however, the light curves of each system exhibited the eclipsing contact binary features. It was assumed that the bolometric albedo was $A_1=A_2=0.5$ (\citealt{1969AcA....19..245R}) and that the gravity-darkening coefficients were $g_1=g_2=0.32$ (\citealt{1967ZA.....65...89L}).
The PHOEBE code provided the limb darkening coefficients as a free parameter, and the stellar atmosphere was modeled using the \cite{2004A&A...419..725C} study.

We set the effective temperature of the hotter stars using the temperature reported by Gaia DR3\footnote{\url{https://gea.esac.esa.int/archive/}}. The effective temperature set on the hotter star is close to the value obtained by the orbital period-effective temperature ($P-T_1$) relationship presented by the \cite{2021ApJS..254...10L} study. Comparing the final effective temperatures of the hotter stars to the Gaia DR3 value, they are in an acceptable range (\citealt{poro2025mnras}).

\begin{figure*}
    \centering
    \includegraphics[width=\textwidth]{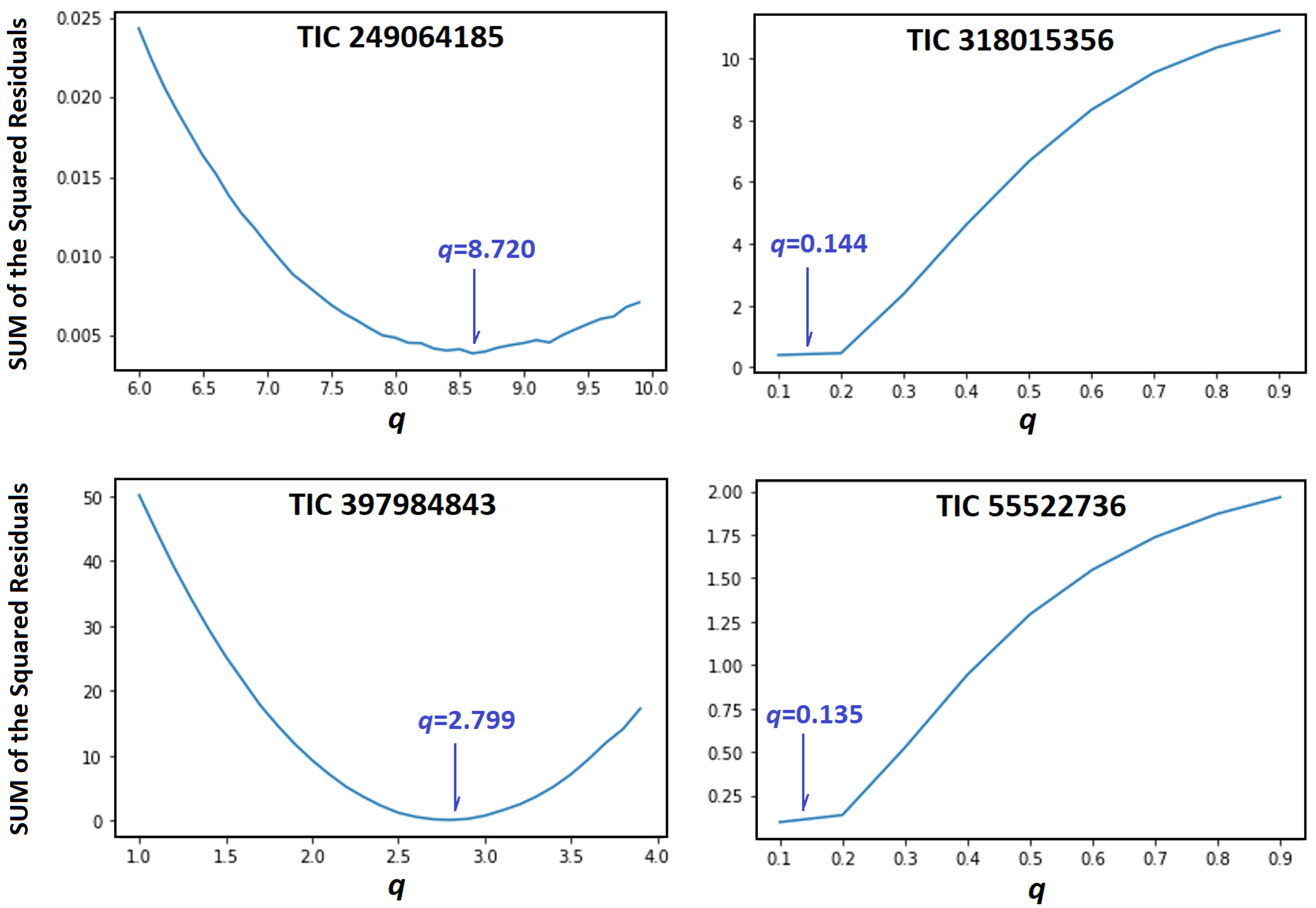}
    \caption{Sum of the squared residuals as a function of the mass ratio.}
    \label{q}
\end{figure*}

We obtained the mass ratio ($q$) of the systems by using the $q$-search in the photometric data.  We searched a range of mass ratios between $q=0.1$ and $q=10$. Then, we shortened the interval and searched again according to the minimum sum of squared residuals. Figure \ref{q} displays the $q$-search results for four systems. The $q$-search curves display a sharp bottom, which allows the determination of an acceptable mass ratio for these total eclipse contact systems (\citealt{2021AJ....162...13L}, \citealt{2024AJ....168..272P}).

The well-known O'Connell effect is indicated by the asymmetry in the brightness of maximums in the light curve of eclipsing binary stars (\citealt{1951MNRAS.111..642O}, \citealt{sriram2017study}). One explanation for this phenomenon could be the existence of starspot(s) due to magnetic activity on the star's surface.

We used PHOEBE's optimization tool to improve the output of the light curve solutions to yield the initial results. The final values of the parameters and their uncertainty were obtained using the MCMC approach based on the emcee package (\citealt{2013PASP..125..306F}). Therefore, five main parameters, including $i$, $q$, $f$, and $T_{1,2}$ were considered for the MCMC modeling process. We selected the Gaussian distribution that adequately encompasses the entire observational light curve and employed 36 walkers and 1500 iterations for target systems. According to the light curve solutions, no target system showed the third body ($l_3$).

\begin{figure*}
    \centering
    \includegraphics[width=\textwidth]{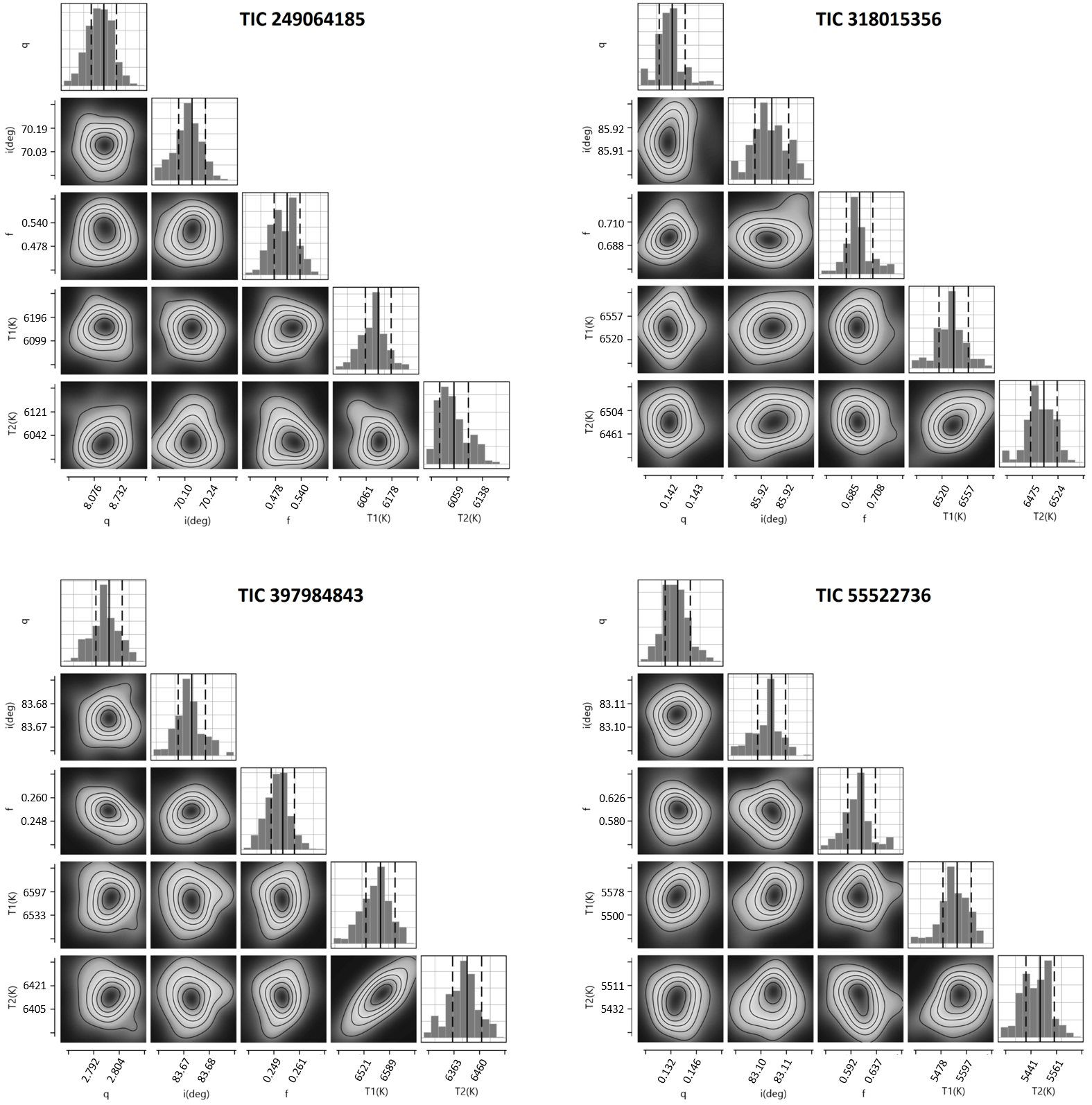}
    \caption{The corner plots of the target systems were determined by MCMC modeling.}
    \label{corner}
\end{figure*}

\begin{figure*}
    \centering
    \includegraphics[scale=0.17]{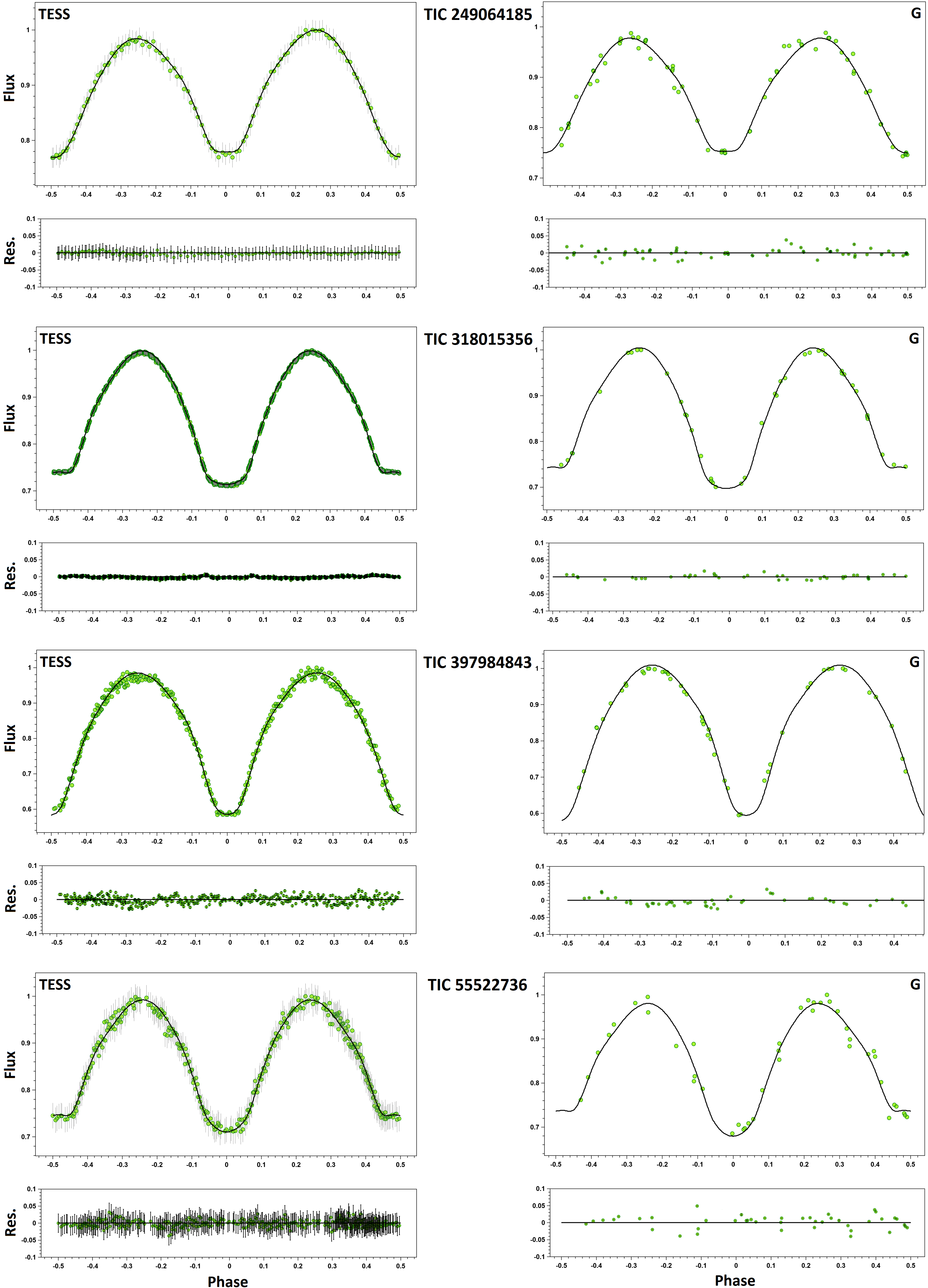}
    \caption{The green dots show the observed light curves of the systems, and the synthetic light curves were generated using the light curve solutions. The right side is the Gaia data, and the left side is the TESS data with their theoretical fits.}
    \label{lc}
\end{figure*}

\begin{figure*}
    \centering
    \includegraphics[width=\textwidth]{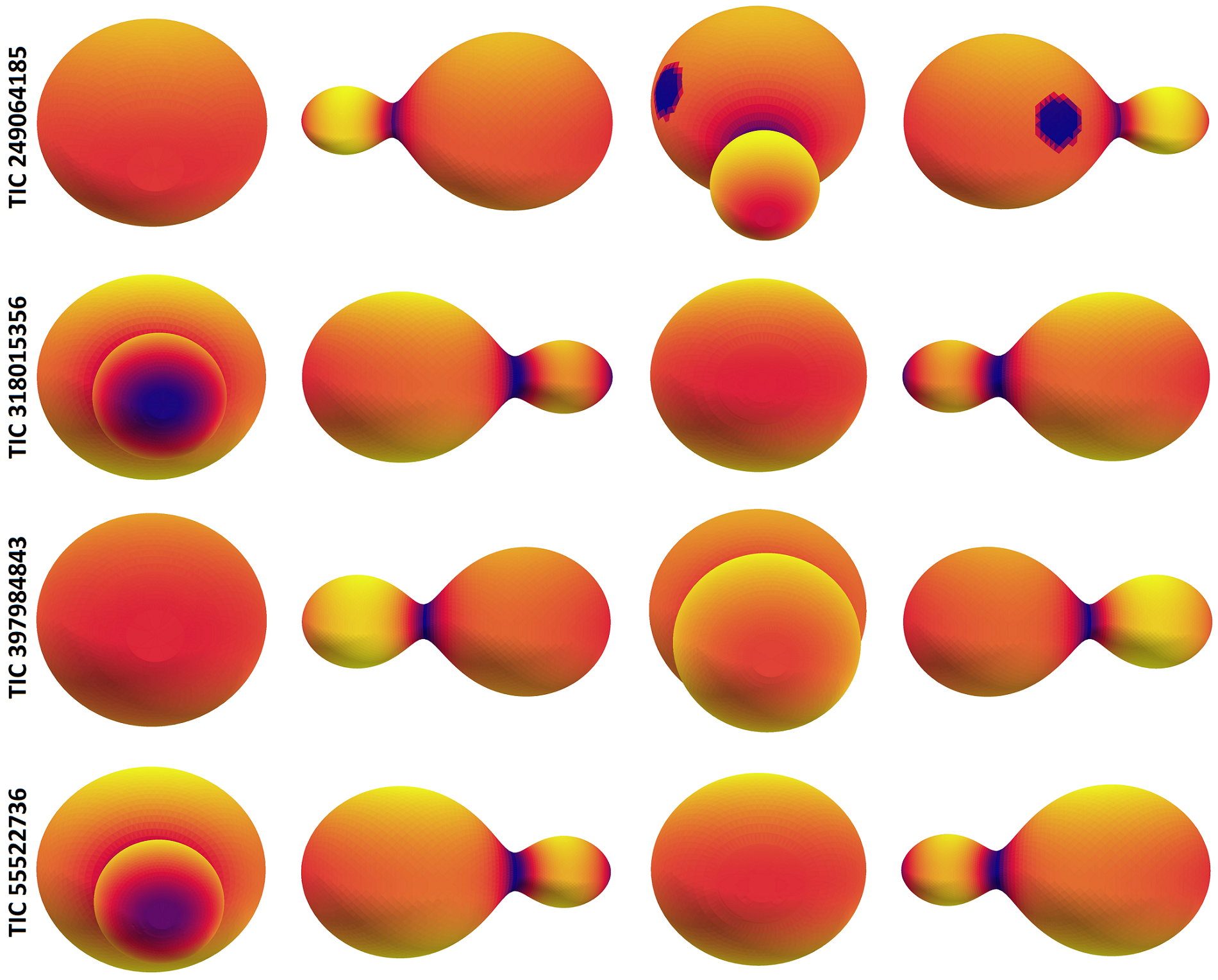}
    \caption{3D view of stars in the four target binary systems in phases 0, 0.25, 0.5, and 0.75.}
    \label{3D}
\end{figure*}

Table \ref{lc-analysis} presents the outcomes of the light curve solutions, Figure \ref{corner} show corner plots of the systems in the MCMC modeling, and Figure \ref{lc} displays the binary systems' observed and final synthetic light curves. The three dimensions (3D) of the binary systems and the starspots on the stars are shown in Figure \ref{3D}. The color in Figure \ref{3D} represents the effective temperature variations on the surface of the stars (\citealt{2023Ap.....66..452P}).

\begin{table*}
\renewcommand\arraystretch{1.8}
\caption{Light curve solutions of the target binary stars.}
\centering
\begin{center}
\footnotesize
\begin{tabular}{c c c c c}
\hline
Parameter & TIC 249064185 & TIC 318015356 & TIC 397984843 & TIC 55522736\\
\hline
$T_{1}$ (K) & $6166_{\rm-(84)}^{+(57)}$ & $6531_{\rm-(24)}^{+(33)}$ & $6569_{\rm-(59)}^{+(47)}$ & $5547_{\rm-(48)}^{+(60)}$\\
$T_{2}$ (K) & $6050_{\rm-(42)}^{+(79)}$ & $6498_{\rm-(26)}^{+(24)}$ & $6417_{\rm-(55)}^{+(45)}$ & $5508_{\rm-(69)}^{+(52)}$\\
$q=M_2/M_1$ & $8.720_{\rm-(450)}^{+(421)}$ & $0.144_{\rm-(1)}^{+(1)}$ & $2.799_{\rm-(8)}^{+(7)}$ & $0.135_{\rm-(7)}^{+(9)}$\\
$i^{\circ}$	& $70.15_{\rm-(11)}^{+(9)}$ & $85.92_{\rm-(1)}^{+(1)}$ & $83.67_{\rm-(1)}^{+(1)}$ & $83.11_{\rm-(1)}^{+(1)}$\\
$f$ & $0.538_{\rm-(45)}^{+(42)}$ & $0.691_{\rm-(11)}^{+(20)}$ & $0.254_{\rm-(7)}^{+(5)}$ & $0.603_{\rm-(37)}^{+(24)}$\\
$A_1=A_2$ & 0.5 & 0.5 & 0.5 & 0.5 \\
$g_1=g_2$ & 0.32 & 0.32 & 0.32 & 0.32 \\
$\Omega_1=\Omega_2$ & 13.259(350) & 2.022(65) & 6.192(157) & 2.009(53)\\
$l_1/l_{tot}$(TESS) & 0.142(4) & 0.839(7) & 0.302(5) & 0.850(5)\\
$l_2/l_{tot}$(TESS) & 0.858(6) & 0.161(5) & 0.698(6) & 0.150(4)\\
$l_1/l_{tot}$($G$) & 0.145(4) & 0.840(7) &0.308(5) & 0.851(5)\\
$l_2/l_{tot}$($G$) & 0.855(6) & 0.160(5) & 0.692(6) & 0.149(4)\\
$r_{(mean)1}$ & 0.235(19) & 0.576(12) & 0.311(17) & 0.577(15)\\
$r_{(mean)2}$ & 0.586(17) & 0.259(15) & 0.488(16) & 0.250(19)\\
Phase shift(TESS) & +0.005(1) & +0.042(2) & -0.011(2) & -0.023(2)\\
\hline										
$Col.^\circ$(spot) & 107 & & & \\
$Long.^\circ$(spot) & 302 & & & \\
$Radius^\circ$(spot) & 13 & & & \\
$T_{spot}/T_{star}$ & 0.89 & & & \\
Component & Secondary & & & \\
\hline
\end{tabular}
\end{center}
\label{lc-analysis}
\end{table*}

\vspace{1cm}
\section{Absolute Parameters}
One of the main objectives of studies on contact binary systems will be to estimate the absolute parameters. Studying the evolution of stellar binary systems involves an appropriate standard of accuracy to estimate absolute parameters such as mass, radius, and luminosity. Using the Gaia DR3 parallax yields more accurate results to estimate absolute parameters (\citealt{2021AJ....162...13L}, \citealt{2024RAA....24b5011P}, \citealt{2024NewA..11002227P}).
In this estimation method, observational parameters such as orbital period $P$ and apparent magnitude $V$, light curve solutions results ($q$, $l_{1,2}/l_{tot}$, $r_{(mean)1,2}$, $T_1$, $T_2$), and Gaia DR3 parallax are used. We used $l_{1,2}/l_{tot}$ of $G$ filter from the light curve solutions. Also, the $V$ magnitude of each system comes from the TIC database for the calculations. The process of calculations and equations used is as follows, respectively (Equations \ref{eq1} to \ref{eq8}).

\begin{equation}\label{eq1}
M_{V(system)}=V-5log(d)+5-A_V
\end{equation}

\begin{equation}\label{eq2}
M_{V(1,2)}-M_{V(tot)}=-2.5log(\frac{l_{(1,2)}}{l_{(tot)}})
\end{equation}

\begin{equation}\label{eq3}
M_{bol}=M_{V}+BC
\end{equation}

\begin{equation}\label{eq4}
M_{bol}-M_{bol_{\odot}}=-2.5log(\frac{L}{L_{\odot}})
\end{equation}

\begin{equation}\label{eq5}
R=(\frac{L}{4\pi \sigma T^{4}})^{1/2}
\end{equation}

\begin{equation}\label{eq6}
a=\frac{R}{r_{mean}}
\end{equation}

\begin{equation}\label{eq7}
\frac{a^3}{G(M_1+M_2)}=\frac{P^2}{4\pi^2}
\end{equation}

\begin{equation}\label{eq8}
g=G_{\odot}(\frac{M}{R^2})
\end{equation}

We calculated the extinction coefficient ($A_V$) value using the 3D dust-map Python package considering the Gaia DR3 reported distance (\citealt{2019ApJ...887...93G}). The \cite{1996ApJ...469..355F} study was utilized to compute the bolometric correction ($BC$) throughout the estimating process.

A reasonable estimating absolute parameters using Gaia DR3 parallax requires a low $A_V$ value (\citealt{2024PASP..136b4201P}). We obtained the absolute parameters using another method to ensure the validity of the estimations since the $A_V$ of the TIC 249064185, TIC 318015356, and TIC 397984843 systems is large (Table \ref{absolute}). So, we used the orbital period and semi-major axis empirical relationship that was updated by the \cite{2024PASP..136b4201P} study (Equation \ref{eqPa}).

\begin{equation}\label{eqPa}
a=(0.372_{\rm-0.114}^{+0.113})+(5.914_{\rm-0.298}^{+0.272})\times P
\end{equation}

Then, we used the mass ratio and Kepler's well-known third law equation to estimate the mass and uncertainty of each star (Equations \ref{eq:M1} and \ref{eq:M2}).

\begin{eqnarray}
M{_1}=\frac{4\pi^2a^3}{GP^2(1+q)}\label{eq:M1}\\
M{_2}=q\times{M{_1}}\label{eq:M2}
\end{eqnarray}

The radius of the stars was estimated using $r_{mean1,2}$ (Equation \ref{eq6}). Then the $L$, $M_{bol}$, and $log(g)$ are possible to calculate using astrophysical equations.

The orbital angular momentum ($J_0$) of the systems was calculated for both estimations method using Equation \ref{eqJ0}, which we utilized from the \cite{2006MNRAS.373.1483E} study:

\begin{equation}\label{eqJ0}
J_0=\frac{q}{(1+q)^2} \sqrt[3] {\frac{G^2}{2\pi}M^5P}
\end{equation}

\noindent where $q$ is the mass ratio, $M$ is the total mass of the system, $P$ is the orbital period, and $G$ is the gravitational constant.

The result of estimating absolute parameters for four systems using Gaia DR3 parallax and $P-a$ methods are listed in Table \ref{absolute}.

\begin{table*}
\renewcommand\arraystretch{1.2}
\caption{Estimated absolute parameters of the systems by two methods.}
\centering
\begin{center}
\footnotesize
\begin{tabular}{c | c c | c c | c c | c c}
\hline
Parameter & \multicolumn{2}{c|}{TIC 249064185} & \multicolumn{2}{c|}{TIC 318015356} & \multicolumn{2}{c|}{TIC 397984843} & \multicolumn{2}{c}{TIC 55522736}\\
& Gaia DR3 & $P-a$ & Gaia DR3 & $P-a$ & Gaia DR3 & $P-a$ & Gaia DR3 & $P-a$\\
\hline
$M_1(M_\odot)$ 	&	 0.13(3)	&	0.16(5)	&	 4.30(26) 	&	1.54(41)	&	 0.77(13) 	&	0.49(13)	&	 3.48(39)	&	1.49(41)	\\
$M_2(M_\odot)$ 	&	 1.16(22) 	&	1.43(50)	&	 0.62(4) 	&	0.22(6)	&	 2.16(37) 	&	1.38(36)	&	 0.47(5)	&	0.20(7)	\\
$R_1(R_\odot)$ 	&	 0.53(6)	&	0.57(10)	&	 2.32(6) 	&	1.63(17)	&	 1.11(11) 	&	0.96(13)	&	 2.06(19)	&	1.54(17)	\\
$R_2(R_\odot)$ 	&	 1.35(16) 	&	1.43(17)	&	 1.03(1) 	&	0.74(11)	&	 1.75(18) 	&	1.50(17)	&	 0.88(7)	&	0.67(11)	\\
$L_1(L_\odot)$ 	&	 0.36(7)	&	0.43(19)	&	 8.82(30) 	&	4.38(1.06)	&	 2.05(34) 	&	1.54(52)	&	 3.62(52)	&	2.04(58)	\\
$L_2(L_\odot)$ 	&	 2.18(44) 	&	2.48(75)	&	 1.68(2) 	&	0.87(29)	&	 4.66(82) 	&	3.45(97)	&	 0.64(8)	&	0.37(16)	\\
$M_{V1}$(mag.)	&	 5.86(19) 	&	5.67(39)	&	 2.37(4) 	&	3.12(23)	&	 3.95(18) 	&	4.25(31)	&	 3.47(14)	&	4.08(26)	\\
$M_{V2}$(mag.)	&	 3.93(21) 	&	4.87(28)	&	 4.17(1) 	&	8.22(31)	&	 3.07(19) 	&	7.18(27)	&	 5.36(12)	&	6.30(37)	\\
$M_{bol1}$(mag.)	&	 5.84(20) 	&	5.65(40)	&	 2.38(4) 	&	3.13(24)	&	 3.96(18) 	&	4.26(32)	&	 3.34(16)	&	3.96(27)	\\
$M_{bol2}$(mag.)	&	 3.90(22) 	&	3.75(29)	&	 4.18(1) 	&	4.88(31)	&	 3.07(19) 	&	3.39(27)	&	 5.23(14)	&	5.81(38)	\\
$log(g)_1$(cgs)	&	 4.11(18) 	&	4.14(3)	&	 4.34(1)	&	4.20(2)	&	 4.24(16) 	&	4.17(1)	&	 4.35(13)	&	4.24(1)	\\
$log(g)_2$(cgs)	&	 4.24(19) 	&	4.28(3)	&	 4.21(1) 	&	4.05(1)	&	 4.29(16) 	&	4.22(1)	&	 4.22(12)	&	4.09(1)	\\
$a(R_\odot)$ 	&	 2.28(14) 	&	2.44(21)	&	 4.00(8) 	&	2.84(23)	&	 3.58(21) 	&	3.08(24)	&	 3.55(13)	&	2.68(23)	\\
$M_{tot}(M_\odot)$ 	&	1.29(25)	&	1.60(55)	&	4.92(29)	&	1.77(47)	&	2.93(51)	&	1.87(49)	&	3.95(44)	&	1.70(48)	\\
$logJ_0(g.cm^2/s)$	&	 51.09(12) 	&	51.25(20)	&	 52.16(5) 	&	51.42(17)	&	 52.05(13) 	&	51.72(17)	&	 51.97(10)	&	51.36(20)	\\
\hline
$A_V(mag.)$	& \multicolumn{2}{c|}{0.504(4)} & \multicolumn{2}{c|}{0.646(2)} & \multicolumn{2}{c|}{1.169(9)} & \multicolumn{2}{c}{0.284(4)}\\
$BC_1$	& \multicolumn{2}{c|}{-0.025(8)} & \multicolumn{2}{c|}{0.008(2)} & \multicolumn{2}{c|}{0.011(4)} & \multicolumn{2}{c}{-0.126(13)}\\
$BC_2$	& \multicolumn{2}{c|}{-0.039(8)} & \multicolumn{2}{c|}{0.006(2)} & \multicolumn{2}{c|}{-0.001(4)} & \multicolumn{2}{c}{-0.135(15)}\\
\hline
\end{tabular}
\end{center}
\label{absolute}
\end{table*}

\vspace{1cm}
\section{Discussion and Conclusion}
We presented the first in-depth light curve solutions and estimations of the absolute parameters of four contact binary stars. We employed the TESS and Gaia survey observations of these binary systems for this investigation.

The light curve solutions show that the lowest effective temperature difference between the two stars is 33 K for TIC 318015356, and the highest is 152 K for TIC 397984843 (Table \ref{ConclusionTable}). The spectral category of the stars of the systems was also presented in Table \ref{ConclusionTable} based on the \cite{2000asqu.book.....C} and \cite{2018MNRAS.479.5491E} studies.

We used the $q$-search method for photometric space-based data for target systems with total eclipses. TIC 249064185, TIC 318015356, and TIC 55522736, with respective mass ratios of $1/q=0.115$, $q=0.144$, and $q=0.135$, are on the border of systems with extremely low mass ratios (\citealt{2024A&A...692L...4L}). Low mass ratio contact binary systems ($q\leqslant0.25$) have been the subject of investigations, and many questions remain (\citealt{2022AJ....164..202L}, \citealt{2024A&A...692L...4L}). Knowledge of the merging process and low-mass ratio limit depends on studies of contact binaries with low-mass ratios.

We employed two methods for estimating absolute parameters. First, the Gaia DR3 parallax was used for this estimation (\citealt{2024NewA..11002227P}, Table \ref{absolute}). The method of using parallax Gaia DR3 is accurate if the restrictions are observed. Using this method requires that the $A_V$ value be low (\citealt{2024PASP..136b4201P}). TIC 249064185, TIC 318015356, and TIC 397984843 have big $A_V$ values, so doubts about the results of this method may arise, although $A_V$ is carefully calculated. On the other hand, the difference in calculating the semi-major axis in the Gaia DR3 parallax method is denoted as $\Delta a$ (\citealt{2024AJ....168..272P}). $\Delta a$ is the separation value from the primary star to the secondary star ($a_1$) and from the secondary star to the primary star ($a_2$). Calculations in the Gaia DR3 parallax method were done from the path of each star separately, and the $a_{1,2}$ values can be different. If $\Delta a$ is less than 0.1, the results are acceptable, and it is also an indication of the accuracy of light curve solutions (\citealt{2024NewA..11002227P}). As shown in Table \ref{ConclusionTable}, $\Delta a$ of the target systems are less than 0.1. However, we also considered the $P-a$ empirical relationship to calculate the absolute parameters.

According to the light curve solutions, all four systems are in total eclipse. Also, light curve solutions and estimation of absolute parameters indicate that the systems are W UMa-type contact binaries. Contact systems can be divided into two A and W subtypes (\citealt{binnendijk1970orbital}). The more massive component is a hotter star in the A-subtype, and if the less massive component has a higher effective temperature, it is classified as a W-subtype. Therefore, the TIC 318015356 and TIC 55522736 systems belong to the A-subtype, while two systems, TIC 249064185 and TIC 397984843, are the W-subtype (Table \ref{ConclusionTable}).

There are three categories for fill-out factors of the contact binary systems: deep ($f\geq 50\%$), medium ($25\% \leq f < 50\%$), and shallow ($f<25\%$) eclipsing contact binary stars (\citealt{2022AJ....164..202L}). According to the results of the fill-out factor in the light curve solutions, there are three target systems of deep type and one system of medium type. We have displayed the fill-out factor category for each target system in Table \ref{ConclusionTable}.

In W UMa-type contact binary systems, understanding the initial mass of two components provides critical insights into their evolutionary processes. We calculated the initial masses of the primary ($M_{1i}$) and secondary ($M_{2i}$) components for the four contact binary systems using the method described by \cite{2013MNRAS.430.2029Y}. The reciprocal mass ratio ($1/{q_i}$) serves as a physical constraint to determine the initial masses of the components. So, the initial mass of the secondary star was estimated by Equation \ref{Mi2} (\citealt{2013MNRAS.430.2029Y}),

\begin{equation}\label{Mi2}
M_{2i}=M_2+\Delta M=M_2+2.50(M_L-M_2-0.07)^{0.64}
\end{equation}

\noindent where $M_2$ is the current mass of the secondary, and $M_L$ is derived from the mass-luminosity relation (Equation \ref{M_L}).

\begin{equation}\label{M_L}
M_L=\left(\frac{L_2}{1.49}\right)^{\frac{1}{4.216}}
\end{equation}

We calculated the initial mass of the primary component using Equation \ref{Mi1},

\begin{equation}\label{Mi1}
M_{1i}=M_1-(\Delta M-M_{\text{lost}})=M_1-\Delta M(1-\gamma)
\end{equation}

\noindent where $M_{\text{lost}}$ represents the mass lost in the system and $\gamma$ is the ratio of $M_{\text{lost}}$ to $\Delta M$ (Equation \ref{Mlost}).

\begin{equation}\label{Mlost}
M_{\text{lost}}=\gamma \times \Delta M
\end{equation}

We set $\gamma=0.664$ based on the \cite{2013MNRAS.430.2029Y} results. Then we used the following equation from the Y 2014 study to determine the ages of the four targets,

\begin{equation}\label{t}
t\approx t_{MS}(M_{2i})+t_{MS}(\overline{M_2}),
\end{equation}

\begin{equation}\label{tMS}
t_{MS}=\frac{10}{(M/M_{\odot})^{4.05}}\times (5.60\times 10^{-3}(\frac{M}{M_{\odot}}+3.993)^{3.16}+0.042)
\end{equation}

\noindent where $\overline{M_2}$ is the average value of $M_{2i}$ and $M_L$. The outcomes, including the initial masses of both stars, the mass loss, and the systems' ages, are presented in Table \ref{ConclusionTable}. We found that these results are consistent with those reported by \cite{2013MNRAS.430.2029Y} and \cite{2014MNRAS.437..185Y}.

\begin{figure*}
    \centering
    \includegraphics[width=\textwidth]{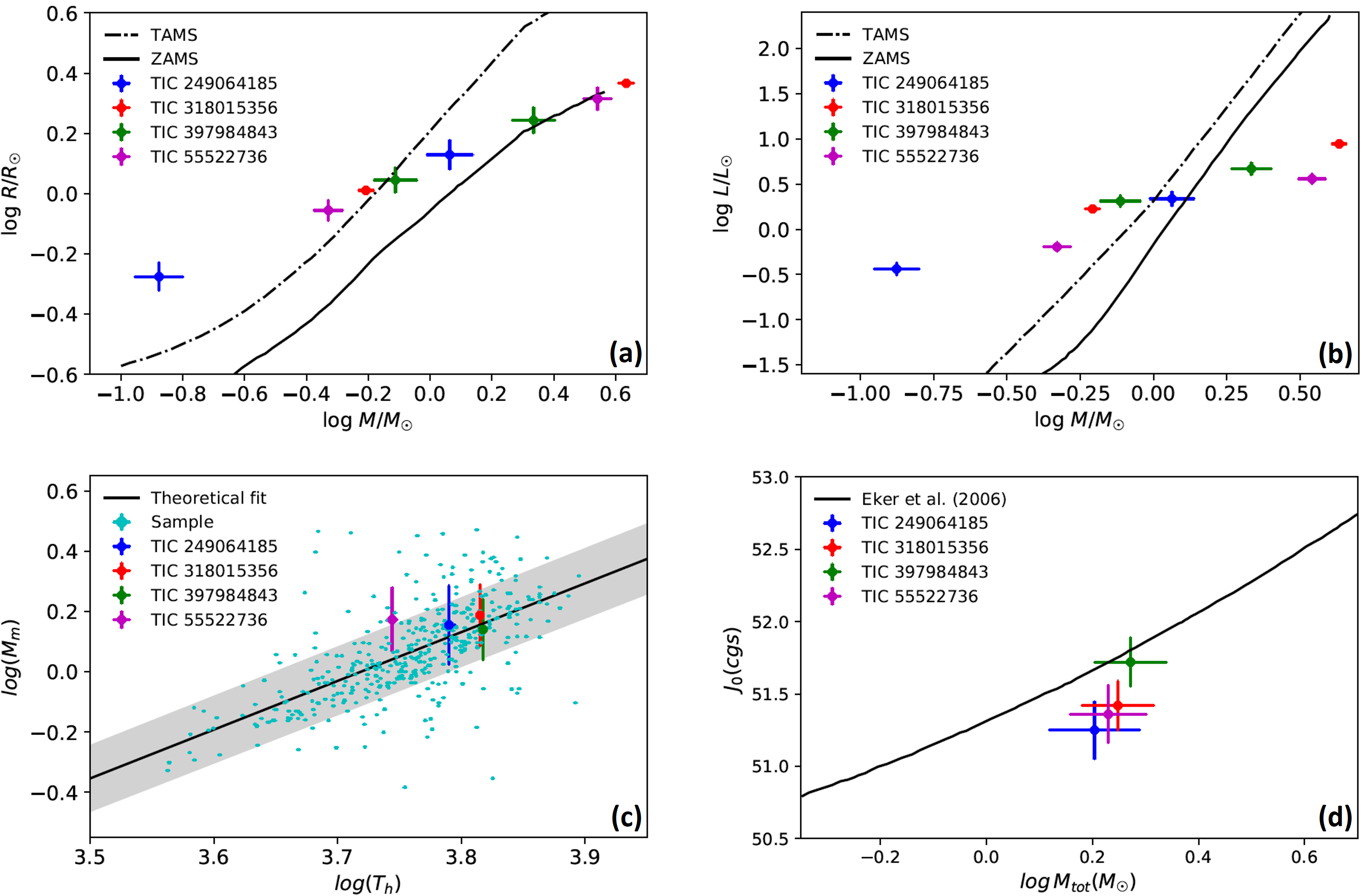}
    \caption{$M-R$, $M-L$, $T_h-M_m$, and $M_{tot}-J_0$ diagrams.}
    \label{relations}
\end{figure*}

Based on the estimation of absolute parameters, We displayed the evolution state of the systems on the Mass-Radius ($M-R$) and Mass-Luminosity ($M-L$) diagrams (Figure \ref{relations}a,b). The figures showed the star positions relative to the Terminal-Age Main Sequence (TAMS) and Zero-Age Main Sequence (ZAMS) lines from the \cite{2000AAS..141..371G} study.

The $T_h-M_m$ relationship was presented by the \cite{2024RAA....24e5001P} study using a sample of 428 contact binary systems. The more massive star was identified as $M_m$, and the hotter component as $T_h$. According to our results, we placed the star's position on the $T_h-M_m$ diagram (Figure \ref{relations}c). There is good agreement between the star locations and the fit of this empirical parameter relationship and its uncertainty.

We showed the location of each system in the $M_{tot}-J_0$ diagram (Figure \ref{relations}d) based on the results in Table \ref{absolute}. The area below the quadratic line in Figure \ref{relations}d is associated with contact binary stars, whereas the area above represents detached systems. Consequently, four target systems are below the quadratic fit and in the contact binary region.

\begin{table*}
\renewcommand\arraystretch{1.2}
\caption{Some conclusions regarding the target systems.}
\centering
\begin{center}
\footnotesize
\begin{tabular}{c c c c c c c c c c}
\hline
System & $\Delta T$ (K) & Sp. & $\Delta a(R_\odot)$ & Subtype & $f_{cat.}$ & $M_{1i}$ ($M_{\odot}$) & $M_{2i}$ ($M_{\odot}$) & $M_{lost}$ ($M_{\odot}$) & $t_{system}$(Gyr)\\
\hline
TIC 249064185 & 116 & F8-G0 & 0.04 & W & Deep & 0.88 & 1.79 & 1.09 & 5.60\\
TIC 318015356 & 33 & F3-F5 & 0.08 & A & Deep & 0.94 & 2.00 & 1.18 & 3.76\\
TIC 397984843 & 152 & F3-F5 & 0.02 & W & Medium & 0.88 & 1.99 & 0.99 & 3.48\\
TIC 55522736 & 39 & G8-G8 & 0.06 & A & Deep & 0.99 & 1.70 & 0.99 & 6.62\\
\hline
\end{tabular}
\end{center}
\label{ConclusionTable}
\end{table*}

\vspace{0.5cm}
\section*{Acknowledgments}
We have made use of data from the European Space Agency (ESA) mission Gaia (\url{http://www.cosmos.esa.int/gaia}), processed by the Gaia Data Processing and Analysis Consortium (DPAC, \url{https://www.cosmos.esa.int/web/gaia/dpac/consortium}). Funding for the DPAC has been provided by national institutions, in particular, the institutions participating in the Gaia Multilateral Agreement.
This work includes data from the TESS mission observations. The NASA Explorer Program funds the TESS mission.
This study uses information from the ASAS-SN variable stars database. ASAS-SN was founded by the Gordon and Betty Moore Foundation through grants GBMF5490 and GBMF10501 to Ohio State University and the Alfred P. Sloan Foundation grant G-2021-14192.
We thank Esfandiar Jahangiri for helping us in the study.

\vspace{0.5cm}
\section*{ORCID iDs}
\noindent Atila Poro: 0000-0002-0196-9732\\
Razieh Aliakbari: 0009-0007-8508-2357\\
Hossein Azarara: 0009-0003-2631-6329\\
Asma Ababafi: 0009-0004-8579-7692\\
Sadegh Nasirian: 0009-0001-8140-1505\\

\bibliography{References}{}

\begin{thebibliography}{}
\expandafter\ifx\csname natexlab\endcsname\relax\def\natexlab#1{#1}\fi
\providecommand{\url}[1]{\href{#1}{#1}}
\providecommand{\dodoi}[1]{doi:~\href{http://doi.org/#1}{\nolinkurl{#1}}}
\providecommand{\doeprint}[1]{\href{http://ascl.net/#1}{\nolinkurl{http://ascl.net/#1}}}
\providecommand{\doarXiv}[1]{\href{https://arxiv.org/abs/#1}{\nolinkurl{https://arxiv.org/abs/#1}}}

\bibitem[{Binnendijk(1970)}]{binnendijk1970orbital}
Binnendijk, L. 1970, vistas in Astronomy, 12, 217

\bibitem[{{Borucki} {et~al.}(2010){Borucki}, {Koch}, {Basri}, {Batalha}, {Brown}, {Caldwell}, {Caldwell}, {Christensen-Dalsgaard}, {Cochran}, {DeVore}, {Dunham}, {Dupree}, {Gautier}, {Geary}, {Gilliland}, {Gould}, {Howell}, {Jenkins}, {Kjeldsen}, {Kondo}, {Latham}, {Lissauer}, {Marcy}, {Meibom}, {Monet}, {Morrison}, {Sasselov}, \& {Tarter}}]{2010AAS...21510101B}
{Borucki}, W.~J., {Koch}, D., {Basri}, G., {et~al.} 2010, in AAS Meeting Abstracts, Vol. 215, AAS Meeting Abstracts \#215, 101.01

\bibitem[{{Castelli} \& {Kurucz}(2004)}]{2004A&A...419..725C}
{Castelli}, F., \& {Kurucz}, R.~L. 2004, \aap, 419, 725, \dodoi{10.1051/0004-6361:20040079}

\bibitem[{{Chen} {et~al.}(2020){Chen}, {Wang}, {Deng}, {de Grijs}, {Yang}, \& {Tian}}]{2020ApJS..249...18C}
{Chen}, X., {Wang}, S., {Deng}, L., {et~al.} 2020, \apjs, 249, 18, \dodoi{10.3847/1538-4365/ab9cae}

\bibitem[{{Conroy} {et~al.}(2020){Conroy}, {Kochoska}, {Hey}, {Pablo}, {Hambleton}, {Jones}, {Giammarco}, {Abdul-Masih}, \& {Pr{\v{s}}a}}]{2020ApJS..250...34C}
{Conroy}, K.~E., {Kochoska}, A., {Hey}, D., {et~al.} 2020, \apjs, 250, 34, \dodoi{10.3847/1538-4365/abb4e2}

\bibitem[{{Cox}(2000)}]{2000asqu.book.....C}
{Cox}, A.~N. 2000, {Allen's astrophysical quantities} (AN Cox ed.)

\bibitem[{{Eker} {et~al.}(2018){Eker}, {Bak{\i}{\c{s}}}, {Bilir}, \& et~al.}]{2018MNRAS.479.5491E}
{Eker}, Z., {Bak{\i}{\c{s}}}, V., {Bilir}, S., \& et~al. 2018, MNRAS, 479, 5491, \dodoi{10.1093/mnras/sty1834}

\bibitem[{{Eker} {et~al.}(2006){Eker}, {Demircan}, {Bilir}, \& {Karata{\c{s}}}}]{2006MNRAS.373.1483E}
{Eker}, Z., {Demircan}, O., {Bilir}, S., \& {Karata{\c{s}}}, Y. 2006, \mnras, 373, 1483, \dodoi{10.1111/j.1365-2966.2006.11073.x}

\bibitem[{{Flower}(1996)}]{1996ApJ...469..355F}
{Flower}, P.~J. 1996, \apj, 469, 355, \dodoi{10.1086/177785}

\bibitem[{{Foreman-Mackey} {et~al.}(2013){Foreman-Mackey}, {Hogg}, {Lang}, \& {Goodman}}]{2013PASP..125..306F}
{Foreman-Mackey}, D., {Hogg}, D.~W., {Lang}, D., \& {Goodman}, J. 2013, \pasp, 125, 306, \dodoi{10.1086/670067}

\bibitem[{{Girardi} {et~al.}(2000){Girardi}, {Bressan}, {Bertelli}, \& {Chiosi}}]{2000AAS..141..371G}
{Girardi}, L., {Bressan}, A., {Bertelli}, G., \& {Chiosi}, C. 2000, AAPs, 141, 371, \dodoi{10.1051/aas:2000126}

\bibitem[{{Green} {et~al.}(2019){Green}, {Schlafly}, {Zucker}, {Speagle}, \& {Finkbeiner}}]{2019ApJ...887...93G}
{Green}, G.~M., {Schlafly}, E., {Zucker}, C., {Speagle}, J.~S., \& {Finkbeiner}, D. 2019, \apj, 887, 93, \dodoi{10.3847/1538-4357/ab5362}

\bibitem[{{Heinze} {et~al.}(2018){Heinze}, {Tonry}, {Denneau}, {Flewelling}, {Stalder}, {Rest}, {Smith}, {Smartt}, \& {Weiland}}]{2018AJ....156..241H}
{Heinze}, A.~N., {Tonry}, J.~L., {Denneau}, L., {et~al.} 2018, \aj, 156, 241, \dodoi{10.3847/1538-3881/aae47f}

\bibitem[{{Henden} {et~al.}(2015){Henden}, {Levine}, {Terrell}, \& {Welch}}]{2015AAS...22533616H}
{Henden}, A.~A., {Levine}, S., {Terrell}, D., \& {Welch}, D.~L. 2015, in American Astronomical Society Meeting Abstracts, Vol. 225, American Astronomical Society Meeting Abstracts \#225, 336.16

\bibitem[{{Jayasinghe} {et~al.}(2018){Jayasinghe}, {Kochanek}, {Stanek}, {Shappee}, {Holoien}, {Thompson}, {Prieto}, {Dong}, {Pawlak}, {Shields}, {Pojmanski}, {Otero}, {Britt}, \& {Will}}]{2018MNRAS.477.3145J}
{Jayasinghe}, T., {Kochanek}, C.~S., {Stanek}, K.~Z., {et~al.} 2018, \mnras, 477, 3145, \dodoi{10.1093/mnras/sty838}

\bibitem[{{Jenkins} {et~al.}(2016){Jenkins}, {Twicken}, {McCauliff}, {Campbell}, {Sanderfer}, {Lung}, {Mansouri-Samani}, {Girouard}, {Tenenbaum}, {Klaus}, {Smith}, {Caldwell}, {Chacon}, {Henze}, {Heiges}, {Latham}, {Morgan}, {Swade}, {Rinehart}, \& {Vanderspek}}]{2016SPIE.9913E..3EJ}
{Jenkins}, J.~M., {Twicken}, J.~D., {McCauliff}, S., {et~al.} 2016, in Society of Photo-Optical Instrumentation Engineers (SPIE) Conference Series, Vol. 9913, Software and Cyberinfrastructure for Astronomy IV, ed. G.~{Chiozzi} \& J.~C. {Guzman}, 99133E, \dodoi{10.1117/12.2233418}

\bibitem[{{Kallrath} \& {Milone}(1999)}]{1999ebs..conf.....K}
{Kallrath}, J., \& {Milone}, E.~F., eds. 1999, {Eclipsing binary stars : modeling and analysis}

\bibitem[{{Kopal}(1959)}]{1959cbs..book.....K}
{Kopal}, Z. 1959, {Close binary systems} (The international astrophysics series)

\bibitem[{{Latkovi{\'c}} {et~al.}(2021){Latkovi{\'c}}, {{\v{C}}eki}, \& {Lazarevi{\'c}}}]{2021ApJS..254...10L}
{Latkovi{\'c}}, O., {{\v{C}}eki}, A., \& {Lazarevi{\'c}}, S. 2021, \apjs, 254, 10, \dodoi{10.3847/1538-4365/abeb23}

\bibitem[{{Li} {et~al.}(2022){Li}, {Gao}, {Liu}, {Gao}, {Li}, {Chen}, \& {Sun}}]{2022AJ....164..202L}
{Li}, K., {Gao}, X., {Liu}, X.-Y., {et~al.} 2022, \aj, 164, 202, \dodoi{10.3847/1538-3881/ac8ff2}

\bibitem[{{Li} {et~al.}(2021){Li}, {Xia}, {Kim}, {Gao}, {Hu}, {Guo}, {Gao}, {Chen}, \& {Guo}}]{2021AJ....162...13L}
{Li}, K., {Xia}, Q.-Q., {Kim}, C.-H., {et~al.} 2021, \aj, 162, 13, \dodoi{10.3847/1538-3881/abfc53}

\bibitem[{{Li} {et~al.}(2024){Li}, {Gao}, {Guo}, {Gao}, {Chen}, {Wang}, {Xin}, {Han}, {Kim}, \& {Jeong}}]{2024A&A...692L...4L}
{Li}, K., {Gao}, X., {Guo}, D.-F., {et~al.} 2024, \aap, 692, L4, \dodoi{10.1051/0004-6361/202451947}

\bibitem[{{Loukaidou} {et~al.}(2022){Loukaidou}, {Gazeas}, {Palafouta}, {Athanasopoulos}, {Zola}, {Liakos}, {Niarchos}, {Hakala}, {Essam}, \& {Hatzidimitriou}}]{2022MNRAS.514.5528L}
{Loukaidou}, G.~A., {Gazeas}, K.~D., {Palafouta}, S., {et~al.} 2022, \mnras, 514, 5528, \dodoi{10.1093/mnras/stab3424}

\bibitem[{{Lucy}(1967)}]{1967ZA.....65...89L}
{Lucy}, L.~B. 1967, \zap, 65, 89

\bibitem[{{Lucy} \& {Wilson}(1979)}]{1979ApJ...231..502L}
{Lucy}, L.~B., \& {Wilson}, R.~E. 1979, \apj, 231, 502, \dodoi{10.1086/157212}

\bibitem[{{Marsh} {et~al.}(2017){Marsh}, {Prince}, {Mahabal}, {Bellm}, {Drake}, \& {Djorgovski}}]{2017MNRAS.465.4678M}
{Marsh}, F.~M., {Prince}, T.~A., {Mahabal}, A.~A., {et~al.} 2017, \mnras, 465, 4678, \dodoi{10.1093/mnras/stw2110}

\bibitem[{{O'Connell}(1951)}]{1951MNRAS.111..642O}
{O'Connell}, D. 1951, MNRAS, 111, 642, \dodoi{10.1093/mnras/111.6.642}

\bibitem[{{Paki} {et~al.}(2023){Paki}, {Baudart}, \& {Poro}}]{2023Ap.....66..452P}
{Paki}, E., {Baudart}, S., \& {Poro}, A. 2023, Astrophysics, 66, 452, \dodoi{10.1007/s10511-024-09802-9}

\bibitem[{{Poro} {et~al.}(2025){Poro}, {Li}, {Paki}, \& et~al.}]{poro2025mnras}
{Poro}, A., {Li}, K., {Paki}, E., \& et~al. 2025, \mnras, \dodoi{10.1093/mnras/staf222}

\bibitem[{{Poro} {et~al.}(2024{\natexlab{a}}){Poro}, {Tanriver}, {Michel}, \& {Paki}}]{2024PASP..136b4201P}
{Poro}, A., {Tanriver}, M., {Michel}, R., \& {Paki}, E. 2024{\natexlab{a}}, \pasp, 136, 024201, \dodoi{10.1088/1538-3873/ad1ed3}

\bibitem[{{Poro} {et~al.}(2024{\natexlab{b}}){Poro}, {Paki}, {Alizadehsabegh}, {Khodadadilori}, {Salehian}, {Hedayatjoo}, {Hashemi}, {Dashti}, \& {Mohammadizadeh}}]{2024RAA....24a5002P}
{Poro}, A., {Paki}, E., {Alizadehsabegh}, A., {et~al.} 2024{\natexlab{b}}, Research in Astronomy and Astrophysics, 24, 015002, \dodoi{10.1088/1674-4527/ad0866}

\bibitem[{{Poro} {et~al.}(2024{\natexlab{c}}){Poro}, {Li}, {Michel}, {Castro}, {Fern{\'a}ndez Laj{\'u}s}, {Wang}, {Coliac}, {Alada{\u{g}}}, {Alizadehsabegh}, \& {Alicavus}}]{2024AJ....168..272P}
{Poro}, A., {Li}, K., {Michel}, R., {et~al.} 2024{\natexlab{c}}, \aj, 168, 272, \dodoi{10.3847/1538-3881/ad834510.1134/S1063772908080015}

\bibitem[{{Poro} {et~al.}(2024{\natexlab{d}}){Poro}, {Jafarzadeh}, {Harzandjadidi}, {Madani}, {Bozorgzadeh}, {Jahangiri}, {Sarostad}, {Alizadehsabegh}, {Hadizadeh}, \& {EsmaeiliVakilabadi}}]{2024RAA....24b5011P}
{Poro}, A., {Jafarzadeh}, S.~J., {Harzandjadidi}, R., {et~al.} 2024{\natexlab{d}}, Research in Astronomy and Astrophysics, 24, 025011, \dodoi{10.1088/1674-4527/ad1b0f}

\bibitem[{{Poro} {et~al.}(2024{\natexlab{e}}){Poro}, {Hedayatjoo}, {Nastaran}, {Nourmohammad}, {Azarara}, {AlipourSoudmand}, {AzarinBarzandig}, {Aliakbari}, {Nasirian}, \& {Kahali Poor}}]{2024NewA..11002227P}
{Poro}, A., {Hedayatjoo}, M., {Nastaran}, M., {et~al.} 2024{\natexlab{e}}, \na, 110, 102227, \dodoi{10.1016/j.newast.2024.102227}

\bibitem[{{Poro} {et~al.}(2024{\natexlab{f}}){Poro}, {Baudart}, {Nourmohammad}, {Sabaghpour Arani}, {Farhadi}, {Ranjbar Salehian}, {Sarostad}, {Ranjbaryan Iri Olya}, {Hadizadeh}, \& {Khodaei}}]{2024RAA....24e5001P}
{Poro}, A., {Baudart}, S., {Nourmohammad}, M., {et~al.} 2024{\natexlab{f}}, Research in Astronomy and Astrophysics, 24, 055001, \dodoi{10.1088/1674-4527/ad3a2c}

\bibitem[{{Pr{\v{s}}a} {et~al.}(2016){Pr{\v{s}}a}, {Conroy}, {Horvat}, {Pablo}, {Kochoska}, {Bloemen}, {Giammarco}, {Hambleton}, \& {Degroote}}]{2016ApJS..227...29P}
{Pr{\v{s}}a}, A., {Conroy}, K.~E., {Horvat}, M., {et~al.} 2016, \apjs, 227, 29, \dodoi{10.3847/1538-4365/227/2/29}

\bibitem[{{Ricker} {et~al.}(2015){Ricker}, {Winn}, {Vanderspek}, {Latham}, {Bakos}, {Bean}, {Berta-Thompson}, {Brown}, {Buchhave}, {Butler}, {Butler}, {Chaplin}, {Charbonneau}, {Christensen-Dalsgaard}, {Clampin}, {Deming}, {Doty}, {De Lee}, {Dressing}, {Dunham}, {Endl}, {Fressin}, {Ge}, {Henning}, {Holman}, {Howard}, {Ida}, {Jenkins}, {Jernigan}, {Johnson}, {Kaltenegger}, {Kawai}, {Kjeldsen}, {Laughlin}, {Levine}, {Lin}, {Lissauer}, {MacQueen}, {Marcy}, {McCullough}, {Morton}, {Narita}, {Paegert}, {Palle}, {Pepe}, {Pepper}, {Quirrenbach}, {Rinehart}, {Sasselov}, {Sato}, {Seager}, {Sozzetti}, {Stassun}, {Sullivan}, {Szentgyorgyi}, {Torres}, {Udry}, \& {Villasenor}}]{2015JATIS...1a4003R}
{Ricker}, G.~R., {Winn}, J.~N., {Vanderspek}, R., {et~al.} 2015, Journal of Astronomical Telescopes, Instruments, and Systems, 1, 014003, \dodoi{10.1117/1.JATIS.1.1.014003}

\bibitem[{{Ruci{\'n}ski}(1969)}]{1969AcA....19..245R}
{Ruci{\'n}ski}, S.~M. 1969, \actaa, 19, 245

\bibitem[{{Samus'} {et~al.}(2017){Samus'}, {Kazarovets}, {Durlevich}, {Kireeva}, \& {Pastukhova}}]{2017ARep...61...80S}
{Samus'}, N.~N., {Kazarovets}, E.~V., {Durlevich}, O.~V., {Kireeva}, N.~N., \& {Pastukhova}, E.~N. 2017, Astronomy Reports, 61, 80, \dodoi{10.1134/S1063772917010085}

\bibitem[{Sriram {et~al.}(2017)Sriram, Malu, Choi, \& Rao}]{sriram2017study}
Sriram, K., Malu, S., Choi, C., \& Rao, P.~V. 2017, The Astronomical Journal, 153, 231

\bibitem[{{Taylor}(2005)}]{2005ASPC..347...29T}
{Taylor}, M.~B. 2005, in Astronomical Society of the Pacific Conference Series, Vol. 347, Astronomical Data Analysis Software and Systems XIV, ed. P.~{Shopbell}, M.~{Britton}, \& R.~{Ebert}, 29

\bibitem[{{Y{\i}ld{\i}z}(2014)}]{2014MNRAS.437..185Y}
{Y{\i}ld{\i}z}, M. 2014, \mnras, 437, 185, \dodoi{10.1093/mnras/stt1874}

\bibitem[{{Yildiz} \& {Do{\u{g}}an}(2013)}]{2013MNRAS.430.2029Y}
{Yildiz}, M., \& {Do{\u{g}}an}, T. 2013, \mnras, 430, 2029, \dodoi{10.1093/mnras/stt028}

\bibitem[{{Zhang} \& {Qian}(2020)}]{2020MNRAS.497.3493Z}
{Zhang}, X.-D., \& {Qian}, S.-B. 2020, \mnras, 497, 3493, \dodoi{10.1093/mnras/staa2166}

\end{thebibliography}
\bibliographystyle{aasjournal}

\end{document}